\documentclass[12pt]{iopart}
\usepackage{iopams} 
\usepackage{bm}
\usepackage{graphicx}
\begin{document}

\title{Transients from initial conditions based on
Lagrangian perturbation theory\\
in $N$-body simulations}

\author{Takayuki Tatekawa$^{\dag,\ddag}$ and Shuntaro Mizuno$^{\|}$}

\address{\dag The center for Continuing Professional Development,
 Kogakuin University, 1-24-2 Nishi-shinjuku,
 Shinjuku, Tokyo 163-8677, JAPAN}
\address{\ddag Advanced Research Institute for Science and Engineering,
Waseda University, 3-4-1 Okubo, Shinjuku,
Tokyo 169-8555, JAPAN}
\address{$\|$ Research Center for the Early Universe (RESCEU),
 School of Science, University of Tokyo,
 7-3-1 Hongo, Bunkyo, Tokyo 113-0033, JAPAN}

\begin{abstract}
We explore the initial conditions for cosmological
N-body simulations suitable for calculating 
the skewness and kurtosis of the density field.
In general, the initial conditions based
on the perturbation theory (PT) provide incorrect
second-order and higher-order growth.
These errors implied by the use of the perturbation theory
to set up the initial conditions in N-body simulations
are called transients.
Unless these transients are completely suppressed 
compared with the dominant growing mode,
we can not reproduce the correct evolution 
of cumulants with orders higher than two,
even though there is no problem with the numerical scheme.
We investigate the impact of transients on the
observable statistical quantities by performing 
$N$-body simulations 
with initial conditions based
on Lagrangian perturbation theory (LPT). 
We show that the effects of 
transients on the kurtosis from the initial conditions, 
based on second-order Lagrangian perturbation theory
(2LPT)
have almost disappeared by $z\sim5$, 
as long as the initial conditions are set at $z > 30$.
This means that for practical purposes, 
the initial conditions based on 2LPT
are accurate enough for numerical calculations of
skewness and kurtosis.

\end{abstract}

\maketitle


\section{Introduction}
 \label{sec:Intro}

The evolution of the large-scale structure in the Universe
is one of the most important topic in astrophysics. 
The standard scenario for the structure formation is 
that the primordial density
fluctuation grows through its gravitational instability
\cite{Peebles,Peacock,Liddle,Coles,Sahni-Coles}. 
Even though the perturbative descriptions are possible
when the density fluctuation is small enough, 
in order to follow the distribution far into the
nonlinear region, we must, inevitably, rely on the
numerical calculations, named, $N$-body simulations
\cite{P3M}.

For $N$-body simulations, there is a problem about
how to set up the initial conditions. Even though 
the naive expectation is that it is better to start
simulations as early as possible, like recombination era,
it is well known that this is 
extremely difficult numerically. 
When we start $N$-body simulations at an early era,
the initial condition can be overwhelmed by sources
of noise, such as the numerical roundoff error or 
the shot noise
of the particles. 
Therefore, in the standard scheme~\cite{Bertschnger1998},
we must use the perturbative approach. 
Until the fluctuations come into the 
quasi-nonlinear regime
and stable numerical calculation become possible.

Even though there has been much progress 
about $N$-body code with how to solve 
the non-linear structure in the universe,
the method to set up the initial conditions into
these codes has been almost the same from 
the first works of this field
\cite{Klypin83,Efstathiou:1985re}.
In many cases,  the Zel'dovich approximation 
(ZA) \cite{ZA}, i.e., the first-order approximation
of Lagrangian perturbation theory (LPT)
have been applied for the initial conditions
of $N$-body simulations for a long time.
(For reviews of LPT, see for example 
\cite{Bernardeau02,Tatekawa04R}).
Even though the ZA describe the growth
of the density fluctuation much better
than the Eulerian linear theory, 
it does not reproduce the higher order statistics, 
like skewness and kurtosis
with very poor accuracy
\cite{Juszkiewicz:1993uw,Juszkiewicz:1993hm,Baugh:1995hv},
because the acceleration is always parallel to the
peculiar velocity. In other words, the acceleration
does not reflect the exact clustering 
in the ZA.

Therefore, $N$-body simulations with 
the ZA initial conditions
fail to pick up the correct second-
and higher-order growing modes hence failing to reproduce
the correct statistical properties of the density
fluctuation until very late times
\cite{Scoccimarro:1997gr,Valageas:2001}.

For this problem, recently, 
Crocce, Pueblas, and Scoccimarro~\cite{Crocce2006}
proposed the improvement by adopting different initial
conditions. Basically, their initial conditions
are based on the approximations valid up to 
second-order Lagrangian perturbation theory (2LPT)
which reproduce exact value of the skewness 
in the weakly nonlinear region
~\cite{Bouchet92,Buchert93}.
With these initial conditions, they calculate 
the statistical quantities and show the effects of 
transients related with 2LPT initial conditions 
decrease much faster than the ones related with ZA
initial conditions,
that is, the transients with 2LPT initial conditions 
are less harmful than ones with ZA
initial conditions.

However, there still exist transients related with 2LPT
initial conditions which prevent to reproduce
the exact value of higher order statistical quantities 
like the kurtosis,
and there is no guarantee that 2LPT initial conditions
are accurate enough for these quantities.

Therefore, as a natural extension of \cite{Crocce2006},
we examine the impact of transients from initial conditions
based on 2LPT in $N$-body simulation.
In this paper, in addition to the ZA and 2LPT, 
we also calculate
the non-Gaussianity with the initial conditions based on
third-order Lagrangian perturbation theory (3LPT)
which reproduce exact
value of the kurtosis in the weakly nonlinear region
\cite{Munshi:1994zb,Buchert94,Bouchet95,Catelan95,Sahni:1995rr}.

This paper is organized as follows.
In Sec.~\ref{sec:NonL}, we present Lagrangian perturbative
solutions valid up to the third-order and briefly explain the
origin of transients. Then, after introducing
the statistical quantities of interest
in this paper in Sec.~\ref{sec:Statistics}, we show
the methods and the results of numerical simulations 
in Sec.~\ref{sec:NumRes}. For these results,
we provide alternative interpretation based on simpler
model  in Sec.~\ref{sec:Nonlinear-effect}.
Sec.~\ref{sec:Summary} is devoted to conclusions.


\section{Lagrangian perturbations} 
\label{sec:NonL}

In this section, we briefly summarize Lagrangian
perturbation theory  (LPT)
and obtain the approximations valid up to the third-order.
We also point out explicitly why transients appear
in the perturbative approach.

Before considering the perturbation, we present the
background cosmic expansion which determines
the motion of cosmic fluid in Newtonian cosmology.
We consider $\Lambda$CDM model in which the Friedmann
equations are given as
\begin{eqnarray}
H^2 &=& \frac{8 \pi G}{3} 
\rho_b + \frac{\Lambda}{3} \,, \\
\frac{1}{a} \frac{d^2 a}{dt^2} &=& - \frac{4\pi G}{3} \rho_b
 + \frac{\Lambda}{3} \,,
\end{eqnarray}
with a background density $\rho_b$ of pressureless fluid 
and a cosmological constant $\Lambda$. 

Even though we take account of $\Lambda$ for the numerical
calculations, we sometimes consider the case 
with $\Lambda=0$, so-called Einstein-de Sitter (E-dS)
model, for a concrete solution  
in this section and the next section.
This can be justified since the effect of $\Lambda$
is negligible at the time we set initial conditions
for cosmological $N$- body simulations.
For simplicity, in this paper, we do not consider 
back-reaction from the motion of the matter to 
cosmic expansion.

Next, we consider the perturbations. 
In this paper, we consider the Lagrangian perturbation
in which solutions for cosmic fluid 
are already derived by several people
\cite{ZA,Bouchet92,Buchert93,Buchert94,Bouchet95,Catelan95}.
For this purpose, it is necessary to define the comoving Lagrangian
coordinates $\bm{q}$ in terms of the comoving Eulerian coordinates
$\bm{x}$ as:
\begin{equation} \label{eqn:x=q+s}
\bm{q} = \bm{x} + \bm{S} (\bm{x},t) \,,
\end{equation}
where $\bm{S}$ is the displacement vector which denotes the deviation from
the homogeneous distribution.

It is worth noting that it is not 
the density contrast $\delta$ but 
the displacement vector $\bm{S}$ that is regarded as
a perturbative quantity in LPT.

In the Lagrangian coordinates, from the continuous equation,
we can express the density contrast exactly as
\begin{equation} 
\label{eqn:L-exactrho}
\delta = J^{-1}-1\,,
\end{equation}
where $J$ is the determinant of the Jacobian of the mapping
between $\bm{q}$ and $\bm{x}$: 
$\partial \bm{x} / \partial \bm{q}$.
From Eq.~(\ref{eqn:L-exactrho}), we can see 
that the break down of the perturbation
with respect to $\delta$ does not necessarily
mean that to $\bm{S}$, which is the strong reason
for considering the Lagrangian picture.
In other words, although the density
diverges or the fluid forms Zel'dovich pancake,
the perturbation does not diverge.

From the physical property, 
$\bm{S}$ can be decomposed
to the longitudinal and the transverse modes:
\begin{eqnarray}
&&\bm{S} = \bm{S}^L + \bm{S}^T \,, \\
&&\nabla_q \times \bm{S}^L = 0 \,, \\
&&\nabla_q \cdot \bm{S}^T = 0 \,,
\end{eqnarray}
where $\nabla_q$ means the Lagrangian spatial derivative.

In this paper, since it is well known that
the transverse mode is negligible for pressureless
fluid, we consider only longitudinal mode.
From  Kelvin circulation theorem,
this is quite natural if it is generated only by the action of gravity.
For the discussions with the transverse mode
in LPT, 
see~\cite{transverse}.

Hereafter we mainly follow the description by Catelan~\cite{Catelan95}.
First, we consider the perturbations in Einstein-de Sitter Universe.
In the Lagrangian description, changing the temporal variable
as  $\tau \equiv t^{-1/3}$, the basic equation
for the density contrast  for the pressureless fluid
named Lagrangian Poisson equation is given by
\begin{eqnarray}
\left[(1+\nabla_q \cdot \bm{S} ) \delta_{\alpha \beta}
-S_{\alpha \beta} +S^C _{\alpha \beta}\right]
\ddot{S}_{\beta \alpha} = \alpha (\tau) 
\left[ J(\bm{q}, \tau) -1\right]\,,\nonumber\\
\label{exact_lag_pois}
\end{eqnarray}
where $S_{\alpha \beta} \equiv \partial S_\alpha/ \partial q_\beta$ is
the deformation tensor,
$S^C _{\alpha \beta}$ is the cofactor matrix of
$S_{\alpha \beta}$. Dots denote the differentiation
with respect to $\tau$ and $\alpha (\tau)$ is
a function of $\tau$ which includes the information
of cosmic expansion law. Especially for 
E-dS universe, $\alpha (\tau) = 6 \tau^{-2}$.

We now solve the dynamical equations for the 
displacements $\bm{S}$ according to the following 
Lagrangian perturbative prescription:
\begin{eqnarray}
\bm{S} (\bm{q}, \tau) =  \bm{S}^{(1)} (\bm{q}, \tau)
+  \bm{S}^{(2)} (\bm{q}, \tau)+ \bm{S}^{(3)} (\bm{q}, \tau)
+\cdots\,.
\label{lagrangian_expansion}
\end{eqnarray}
Here $\bm{S}^{(1)}$ corresponds to the
first-order approximation,
$ \bm{S}^{(2)}$ to the second-order approximation,
and so on. Since we consider
only the longitudinal modes,
the perturbative solutions are described with
gradient of scalar functions,
\begin{equation}
\bm{S}^{(n)} \equiv \nabla_q S^{(n)} \,.
\end{equation}

For the pressureless fluid,
it can be shown that the perturbative solutions
are separable into the temporal and the spatial parts,
\begin{eqnarray}
\bm{S}^{(1)} (\bm{q}, \tau) &=& 
D(\tau) \bm{s}^{(1)} (\bm{q}) \,, \\
\bm{S}^{(2)} (\bm{q}, \tau) &=& 
E(\tau) \bm{s}^{(2)} (\bm{q}) \,, \\
\bm{S}^{(3)}  (\bm{q}, \tau) &=& 
F(\tau) \bm{s}^{(3)} (\bm{q}) \,.
\end{eqnarray}
The dynamics of the evolution constrains
in general both the temporal dependence as
described by the functions $D$, $E$, $F$, $\ldots$
and the spatial displacements 
$\bm{s}^{(n)}$.

First we derive the Lagrangian linear perturbative solution
(ZA). Making use of the fact that up to the third-order,
the following expression for the Jacobian determinant
$J$ holds~\cite{Catelan95}:
\begin{eqnarray}
J&&= 1 + \nabla_q \cdot  \bm{S}^{(1)} +  
\nabla_q \cdot  \bm{S}^{(2)} + \frac{1}{2}
\left[(\nabla_q \cdot \bm{S}^{(1)} )^2 -
S_{\alpha \beta} ^{(1)} S_{\beta\alpha} ^{(1)}\right]
\nonumber\\
&&+ \nabla_q \cdot  \bm{S}^{(3)} + 
\left[(\nabla_q \cdot \bm{S}^{(1)} )
(\nabla_q \cdot \bm{S}^{(2)} )  -
S_{\alpha \beta} ^{(1)} S_{\beta\alpha} ^{(2)}\right]
\nonumber\\
&&+\det (S_{\alpha \beta} ^{(1)})\,,
\end{eqnarray}
we can easily find the first-order approximation
truncating Eq.~(\ref{exact_lag_pois}), which 
yields,
\begin{eqnarray}
\ddot{D}(\tau) - \alpha(\tau) D(\tau) =0\,,
\label{linear_lag_pois}
\end{eqnarray}
and no constraint to $\bm{s}^{(1)} (\bm{q})$.

For the E-dS model, by substituting
$\alpha (\tau) = 6 \tau^{-2}$ into 
Eq.~(\ref{linear_lag_pois}), we can obtain
the concrete form of the growing mode
and the decaying mode solution as
\begin{eqnarray}
D_+(t) &=& \left ( \frac{t}{t_0} \right )^{2/3} \,, \\
D_-(t) &=& \left ( \frac{t}{t_0} \right )^{-1} \,.
\end{eqnarray}
In general, we consider only the growing mode
because the decaying mode is suppressed by
$a^{-5/2}$ compared to the growing mode
and soon becomes negligible.

It is worth noting that the analytic
solution for the linear perturbative solution
is obtained even in the case with $\Lambda$.
The perturbative solutions are described as~\cite{Bouchet95}
\begin{eqnarray}
D_+ (t) &=& \frac{h}{2} B_{1/h^2} \left (\frac{5}{6}, \frac{2}{3} \right )
 \,, \label{growth} \\
D_- (t) &=& h \,, \\
h &=& \sqrt{\frac{3}{\Lambda}} \frac{\dot{a}}{a} \,,
\end{eqnarray}
where $B_{1/h^2}$ is incomplete Beta function:
\begin{equation}
B_z (\mu, \nu) \equiv \int_0^z p^{\mu-1} (1-p)^{\nu-1}
 {\rm d} p \,.
\end{equation}
Hereafter we consider growing mode ($D_+$) only and 
re-define $D$ as $D_+$. Even though the impact is not
so significant, various modes appear 
in higher-order perturbation when we consider $D_-$.
For detail, see \cite{Sasaki98}.

Next, we construct the solution valid up to 
second-order Lagrangian perturbation theory (2LPT).
Retaining only the quadratic terms in 
Eq.~(\ref{exact_lag_pois}) and using the first-order
results, the system of equations become as follows:
\begin{eqnarray}
&&\ddot{E}(\tau) - \alpha (\tau) E(\tau)
= - \alpha (\tau) D(\tau)^2\, , 
\label{2nd_lag_pois_temp}\\
&& \nabla_q \cdot  \bm{s}^{(2)} 
= \frac{1}{2}
\left[(\nabla \cdot \bm{s}^{(1)} )^2 -
s_{\alpha, \beta} ^{(1)} s_{\beta, \alpha} ^{(1)}
\right]\,,
\label{2nd_lag_pois_spa}
\end{eqnarray}
where $s_{\alpha, \beta}= \partial s_{\alpha}/ \partial q_{\beta}$.
From Eqs.~(\ref{2nd_lag_pois_temp}) and 
(\ref{2nd_lag_pois_spa}), we see that 
unlike the linear case, the second-order 
approximation constraints both the temporal and the spatial
dependence of the solution.

In the E-dS model, by substituting 
$\alpha (\tau) = 6 \tau^{-2}$ into 
Eq.~(\ref{2nd_lag_pois_temp}), the growing mode solution
can be obtained as
\begin{equation} 
E_+ (t) = -\frac{3}{7} \left ( \frac{t}{t_0} \right )^{4/3}
 = -\frac{3}{7} D(t)^2 \,. 
\end{equation}
In addition to this, the solutions satisfying 
Eq.~(\ref{2nd_lag_pois_temp}) without the source term
like
\begin{equation} 
E_- (t) = c_1 t^{2/3} + c_2 t^{-1}\,,
\end{equation}
where $c_1$ and $c_2$ are constants,
also satisfy Eq.~(\ref{2nd_lag_pois_temp}).

Of course, $E_- (t)$ can be regarded as 
the ``decaying modes'' and after some time become
negligible regardless of the concrete value
of $c_1$ and $c_2$ compared to the growing mode
$E_+ (t)$. 
They are, however, suppressed by only $a^{-1}$, and
if the initial conditions do not use the exact
value for $c_1$ and $c_2$, the error survives
until late time. These nonphysical decaying modes
are well known {\it transients}.

Actually, if the initial conditions for
$N$-body simulations are set by the ZA, 
these are appropriate at only
linear-order and 
are inappropriate at second- and higher-order. 
Therefore, we cannot 
obtain the correct $c_1$ and $c_2$ as long as the initial
conditions are set by the ZA. This is the point we discuss
in this paper.
(For the complete discussions about the origin of 
transients, see e.g. Sec.~2.1 in \cite{Crocce2006}.)

On the other hand, from Eq.~(\ref{2nd_lag_pois_spa})
the spatial part can be described as
\begin{eqnarray} \label{2nd-s}
\bm{s}^{(2)} &=& \frac{1}{2} \left[ \bm{s}^{(1)}
 \left (\nabla_q \cdot \bm{s}^{(1)} \right )
 - \left ( \bm{s}^{(1)} \cdot \nabla_q \right )
 \bm{s}^{(1)} \right ] \nonumber \\
 && + \bm{R}^{(2)} \,, 
\end{eqnarray}
where $\bm{R}^{(2)}$ is a
divergence-free vector such that 
$\nabla_q \times \bm{s}^{(2)} = 0$.

Then, to see if the effect of transients can be
suppressed by considering higher order terms,
we continue to analyze the solution valid up to 
third-order Lagrangian perturbation theory (3LPT).
Inserting the expansion 
Eq.~(\ref{lagrangian_expansion}) into
Lagrangian Poisson equation, and using the results
of the linear order and the second-order,
we can obtain the third-order expression.
It is more convenient to split the third-order
displacement $\bm{S}^{(3)}$ into two parts as follows:
\begin{equation} 
\bm{S}^{(3)} (\bm{q})= 
\bm{S}_a ^{(3)} (\bm{q}) + \bm{S}_b ^{(3)} (\bm{q})\,, 
\end{equation}
where $\bm{S}_a^{(3)}$ is from the cubic interaction
among the linear perturbations and $\bm{S}_b ^{(3)}$
is from the interaction between the linear and 
the second-order perturbations.

Then, the part for $\bm{S}_a ^{(3)}$
is constrained by
\begin{eqnarray} 
&&\ddot{F}_a (\tau) - \alpha(\tau) F_a (\tau)
=-2\alpha(\tau) D(\tau)^3\,,
\label{3rda_lag_pois_temp}\\
&&\nabla_q \cdot \bm{s}_a ^{(3)} = 
\det (s_{\alpha, \beta} ^{(1)})\,.
\label{3rda_lag_pois_spa}
\end{eqnarray} 

In the E-dS model, by substituting 
$\alpha (\tau) = 6 \tau^{-2}$ into 
Eq.~(\ref{3rda_lag_pois_temp}), 
the growing mode solution can be obtained as
\begin{equation} 
{F_a}_+ (t) = -\frac{1}{3} \left ( \frac{t}{t_0} \right )^2
 = -\frac{1}{3} D(t)^3 \,.
\end{equation}
In addition to this, the solutions satisfying
Eq.~(\ref{3rda_lag_pois_temp}) without the source term
like
\begin{equation} 
{F_a}_- (t) = c_3 t^{2/3} + c_4 t^{-1}\,,
\end{equation}
where $c_3$ and $c_4$ are constants,
also satisfy Eq.~(\ref{3rda_lag_pois_temp})
and they serve as transients, too,
unless the initial condition are set appropriately.

However, compared with the growing mode,
these transients are suppressed by $a^{-2}$
and they are less problematic than the 
transients related to the second-order perturbation,
which are suppressed by $a^{-1}$.

From Eq.~(\ref{3rda_lag_pois_spa}) we can obtain
the spatial part of $\bm{S}_a ^{(3)}$ as
\begin{equation} \label{3rd-sa}
s_{a \alpha} ^{(3)} = \frac{1}{3} 
s_{\alpha \beta} ^{(1) C} s_\beta ^{(1)}
+R_{a \alpha} ^{(3)}\,,
\end{equation}
where $\bm{R}_a^{(3)}$ is a divergence-free vector
such that  $\nabla_q \times \bm{s}_a^{(3)} = 0$.

On the other hand, the part $\bm{S}_b^{(3)}$
is constrained by
\begin{eqnarray} 
\ddot{F}_b (\tau) - \alpha(\tau) F_b (\tau)
=-2\alpha(\tau) D(\tau)[E(\tau) -D(\tau)^2]\,,
\label{3rdb_lag_pois_temp}\\
\nabla_q \cdot \bm{s}_b ^{(3)} = 
\left[(\nabla_q \cdot \bm{s}^{(1)} )
(\nabla_q \cdot \bm{s}^{(2)} )  -
s_{\alpha, \beta} ^{(1)} s_{\beta , \alpha} ^{(2)}\right]\,.
\label{3rdb_lag_pois_spa}
\end{eqnarray} 
From the same discussion for the part $\bm{S}_a^{(3)}$,
we can acquire the growing mode solution in 
the E-dS model,
\begin{equation} 
{F_b}_+ (t) = \frac{10}{21} \left ( \frac{t}{t_0} \right )^2
 = \frac{10}{21} D(t)^3 \,,
\end{equation} 
as well as the spatial part of $\bm{S}_b^{(3)}$,
\begin{eqnarray} 
\bm{s}_{b}^{(3)} &=& \frac{1}{4} 
\left [ \bm{s}^{(1)} 
\left (\nabla_q \cdot \bm{s}^{(2)} \right )
 - \left ( \bm{s}^{(1)} \cdot \nabla_q \right ) \bm{s}^{(2)} 
 \right. \nonumber \\
  && \left .
 + \bm{s}^{(2)} \left (\nabla_q \cdot \bm{s}^{(1)} \right )
 - \left ( \bm{s}^{(2)} \cdot \nabla_q \right ) \bm{s}^{(1)} 
 \right ] \nonumber \\
  && + \bm{R}_B^{(3)} \label{3rd-sb} \,,
\end{eqnarray} 
where $\bm{R}_b^{(3)}$ is again a divergence-free vector
such that  $\nabla_q \times \bm{s}_b^{(3)} = 0$.


\section{Statistics} \label{sec:Statistics}

Here, we introduce the statistical 
quantities of the density fields on which 
transients provide the impact at large scales.
For this purpose, a one-point probability distribution function
of the density fluctuation field $P(\delta)$
(PDF of the density fluctuation) which denotes 
the probability of obtaining the value $\delta$
plays an important role.
If $\delta$ is a random Gaussian field, 
the PDF of the density 
fluctuation is determined as
\begin{eqnarray}
P(\delta) = \frac{1}{(2\pi \sigma^2)^{1/2}}
 e^{-\delta^2/2\sigma^2}\,,
\label{Gauss}
\end{eqnarray}
where 
$\sigma^2 \equiv \left < \left (\delta-\left <\delta 
\right> \right )^2 \right > $ 
is the dispersion and $\left <\;\;\right >$ denotes 
the spatial average.

Since linear growing modes are reproduced 
almost exactly by the ZA initial condition, 
the power spectrum, which is the quantity obtained  
by linear perturbation theory, is not affected by transients.
Therefore, the expected tools to detect
transients at large scales are 
the cumulants of the one-point PDF of the density fluctuation
field whose orders are higher than two
which  become nonzero for the distribution
deviating from Gaussian.
In the structure formation, non-Gaussianity are 
generated because of the non-linear dynamics
of the fluctuations, even though $\delta$
is initially treated as a random Gaussian field, 
as a result of the generic prediction of inflationary
scenario.

For this purpose, we concentrate on
the third and fourth order cumulants,
which are defined as 
$<\delta^3>_c \equiv <\delta^3>$,
$<\delta^4>_c \equiv <\delta^4>-3\sigma^4$
display asymmetry and non-Gaussian degree of
``peakiness'', respectively, for a given dispersion
\cite{Peebles,Peacock}. Since
it is known that the scaling 
$<\delta^n>_c \propto \sigma^{2n-2}$ holds
for weakly non-linear regions during 
the gravitational clustering 
\cite{Bernardeau02}
from Gaussian initial conditions,
 we introduce 
the following normalized higher-order statistical
quantities~:
\begin{eqnarray*}
\mbox{skewness} &:& \gamma = \frac{ \left < \delta ^3 \right >_c }
{\sigma^4} \,, \\
\mbox{kurtosis} &:& \eta = 
\frac{\left < \delta^4 \right >_c}{\sigma^6} \,.
\end{eqnarray*}
The merit of adopting these definitions is, as stated above,
that they are constants in weakly nonlinear stage
which are given by Eulerian linear and second-order
perturbation theory \cite{Peebles,Bernardeau02}. 
For example, 
in the E-dS model smoothed with a spherical top-hat
window function,
\begin{equation}
\tilde{W} = \frac{3(\sin x - x \cos x)}{x^3}\,,
\label{tophat_window}
\end{equation}
the skewness and the kurtosis are given by
\begin{eqnarray}
\gamma &=& \frac{34}{7} + y_1 + {\cal{O}}(\sigma^2) \, ,\\
\eta &=& \frac{60712}{1323} + \frac{62}{3}y_1
-\frac{7}{3}y_1^3 + \frac{2}{3}y_2 
+{\cal{O}}(\sigma^2) \, ,
\end{eqnarray}
where
\begin{eqnarray}
y_p \equiv  \frac{d^p \ln \sigma^2(R)}
{d \ln^p R},
\end{eqnarray}
with smoothing scale $R$~\cite{Bernardeau02}.

For this form of the skewness and the kurtosis,
the effects of transients at large scales
from the ZA initial condition is also investigated by
\cite{Scoccimarro:1997gr,Valageas:2001} 
as 
\begin{eqnarray}
\gamma_{\rm tran} &=& - \frac{6}{5a} + \frac{12}{35 a^{7/2}}\,,
\label{transients_ZA_initial_skew} \\
\eta_{\rm tran} &=& -\frac{816}{35a}-\frac{28 y_1}{5a}
+\frac{184}{75a^2}+\frac{1312}{245a^{7/2}}
\nonumber\\
&&+\frac{8 y_1}{5 a^{7/2}}-\frac{1504}{4725a^{9/2}}
+\frac{192}{1225a^7}\,.
\label{transients_ZA_initial_kurt}
\end{eqnarray}

For the initial condition, instead of the ZA, 
if we adopt the one based on 2LPT, transients
in the skewness (\ref{transients_ZA_initial_skew}) 
is expected to be vanished at large scales,
since 2LPT can provide appropriate initial condition
up to second order related with the skewness.
Similarly, for the one based on 3LPT,
 transients
in the kurtosis (\ref{transients_ZA_initial_kurt}) 
is also expected to be vanished at large scales.

Essentially, in the following section, 
we examine the effects of transients 
in the $\Lambda$CDM model relying on
$N$-body simulation from the 
initial conditions based on 2LPT and 3LPT.
Since we investigate numerically, we can 
go into the highly nonlinear region,
in which the analytic estimates of transients
like Eqs.~(\ref{transients_ZA_initial_skew})
and (\ref{transients_ZA_initial_kurt}) 
can not be performed.


\section{Numerical Calculations} \label{sec:NumRes}

In this section, we calculate the statistical quantities
introduced in the previous section in the
$\Lambda$CDM model based on $N$-body simulations. 
For setting up the initial conditions, 
we use COSMICS code~\cite{COSMICS}
which generates primordial Gaussian density field
usually based on the ZA.
We consider the case with those based on 2LPT and 3LPT, too.
COSMICS package consists of 4 applications. GRAFIC generated
Gaussian random density fields (density, velocity, and
particle displacements) on a lattice. Both the velocity
and the displacements are related to each other.

GRAFIC automatically selects the output redshift
by the maximum density fluctuation on a grid 
$\delta_{max}$ for a given set
of cosmological parameters. 
In order to obtain the initial redshift, we adopt the
following cosmological parameters
at the present time ($z=0$) which are given
by WMAP 3-year result~\cite{WMAP}:
\begin{eqnarray}
\Omega_m &=& 0.28 \,, \\
\Omega_{\Lambda} &=& 0.72 \,, \\
H_0 &=& 73~ \mbox{[km/s/Mpc]} \,, \\
\sigma_8 &=& 0.74 \,.
\end{eqnarray}
The averaged relation between the input maximum density
fluctuation and the output redshift is shown 
in Table.~\ref{tab:redshift}.
The initial redshift is set by
the input maximum density fluctuation. Because we
set random Gaussian fluctuation, the initial redshift
is not fixed.

From the initial conditions set up above, we
follow the evolution of the particles based on
N-body simulation.
The numerical algorithm is applied 
by particle-particle particle-mesh
($P^3M$) method~\cite{P3M} which was developed by
Gelb and Bertschinger. The numerical code we use
is written by Bertschinger.
For N-body simulations, we set the parameters as follows:
\begin{eqnarray*}
\mbox{Number of particles} &:& N=128^3 \,, \\
\mbox{Box size} &:& L=128 h^{-1} \mbox{Mpc}
 ~~(\mbox{at}~z=0)  \,, \\
\mbox{Softening length} &:& \varepsilon = 50 h^{-1} \mbox{kpc}
 ~~(\mbox{at}~z=0)  \,.
\end{eqnarray*}

For the simulations, we use 50 samples for an initial
condition. After the calculations,
in order to avoid the divergence of the density
fluctuation in the limit of large $k$, however,
just for a technical reason,
it is necessary to consider the density field 
$\rho_m(\bm{x};R)$
at the position $\bm{x}$ smoothed over the scale $R$,
which is related to the unsmoothed density field
$\rho_m(\bm{x})$ as
\begin{eqnarray}
\rho_m(\bm{x};R) &=& \int d^3 \bm{y} W(|\bm{x}-\bm{y}|;R)
 \rho_m(\bm{y}) \,,
\end{eqnarray}
where we use the top-hat spherical window function
by Eq.~(\ref{tophat_window}).
Throughout this paper, we choose the smoothing scale
$R=2 h^{-1} \mbox{Mpc} ~~(\mbox{at}~z=0)$.

Crocce, Pueblas, and Scoccimarro~\cite{Crocce2006} analyzed
the non-Gaussianity of the density field with 
both the skewness and the kurtosis.
They changed the smoothing scale $R$ and compared 
the non-Gaussianity with two different initial conditions
which are  based on the ZA and 2LPT, respectively.
Then they showed that the difference of
the skewness and the kurtosis
for $R=2 h^{-1}$ Mpc becomes several percent at $z=0$,
when the initial condition is imposed at $z=24$.

Here, in addition to the ZA and 2LPT, we also 
analyze the non-Gaussianity 
with the initial conditions based on 3LPT,
by slightly modifying the COSMICS
so that initial conditions are set up by
not only the ZA but also 2LPT and 3LPT.
Using output files which describe the ZA displacement
from COSMICS, we compute 2LPT and 3LPT displacements
and velocities using Eqs.~(\ref{2nd-s}), (\ref{3rd-sa}),
and (\ref{3rd-sb}). Here we set
$\bm{R}^{(2)}=\bm{R}_A^{(3)}=\bm{R}_B^{(3)}=\bm{0}$.
The vectors $\bm{R}^{(2)}, \bm{R}_A^{(3)}$, and $\bm{R}_B^{(3)}$
were introduced for rotation-free condition. Because we
set these vectors as zero, the rotation-free condition
slightly violates.

The main purpose is to see the 
impacts of transients from 2LPT initial conditions 
since 3LPT initial conditions
are expected to provide exact results
for the kurtosis.
We investigate this also by changing 
the initial time dependence 
which corresponds to the initial maximum 
density fluctuation for a given  Fourier mode. 
We expect this can clarify the range where
N-body simulations from 2LPT initial conditions 
are accurate enough.

\begin{table}
\caption{\label{tab:redshift} The averaged redshift and
its dispersion of the initial condition selected by
the GRAFIC code for a given maximum density fluctuation.
}
\begin{tabular}{lcc}
\hline \hline
$\delta_{max}$ & $\bar{z}$ & $\sigma_z$ \\ \hline
$1.0$ & $22.821$ & $0.868$ \\
$0.5$ & $32.833$ & $1.670$ \\
$0.2$ & $83.806$ & $4.256$ \\ \hline
\end{tabular}
\end{table}

First, we analyze the case in which the initial condition
is imposed after the density fluctuation becomes relatively
large ($\delta_{max} (t_i) =1$).

\begin{figure}[tb]
\centerline{
\includegraphics[height=5cm]{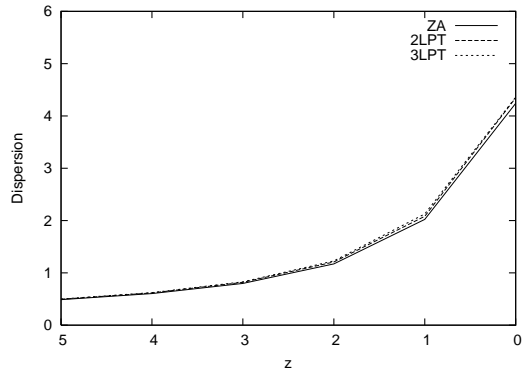}
}
\caption{The dispersion of the density distribution
from N-body simulation ($R = 2 h^{-1}$ Mpc) with different
initial conditions. 
The difference between dispersion with the initial conditions
based on the ZA and 3LPT is less than $4 \%$ at $z=0$.
}
\label{fig:d10-sigma}
\end{figure}

Before analyzing the non-Gaussianity, we consider 
the evolution of the dispersion of the density distribution
(Fig.~\ref{fig:d10-sigma}). We can see that the
ZA is a good approximation initially, but it gradually 
deviates from the one from 2LPT initial condition
because of the nonlinear effects. The final
difference between the dispersions 
with the initial conditions based on the ZA and 2LPT
is less than $4 \%$. There is almost
no difference between the dispersions
in density with the initial
conditions based on 2LPT and 3LPT.
Obviously, this confirms the strength of ZA,
that can reproduce the 
growth of the density fluctuations
well within the quasi-nonlinear regime.

\begin{figure}[tb]
\centerline{
\includegraphics[height=12cm]{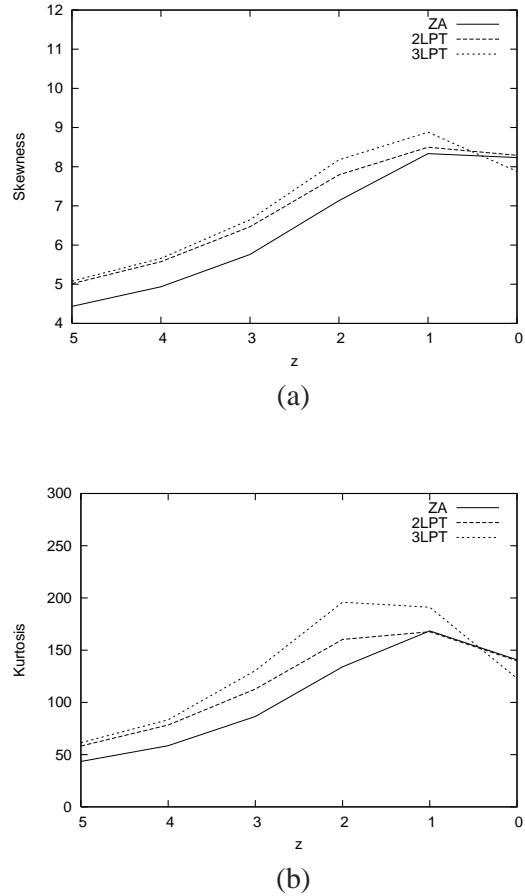}
}
\caption{The non-Gaussianity of the density distribution
from N-body simulation 
($\delta_{max} (t_i)=1.0, R = 2 h^{-1}$ Mpc)
with different initial conditions.
(a) The skewness of the density distribution.
The difference between the skewness with the initial conditions
based on the ZA and 3LPT is less than  
$14 \%$. Furthermore, the skewness with the initial condition
based on 2LPT does not coincide with that based on 3LPT
obviously because of the high nonlinearity of the initial time.
(b) The kurtosis of the density distribution. 
The difference between the kurtosis 
with the initial conditions
based on ZA and 3LPT is less than $40 \%$. 
Furthermore, the kurtosis with the initial condition
based on 2LPT does not coincide with that based on 3LPT
obviously.}
\label{fig:d10-NG}
\end{figure}

However we also see that when it comes to 
the non-Gaussianity which include the skewness
and higher-order moments, the validity of
the ZA is very limited
(Fig.~\ref{fig:d10-NG}).
The difference between the skewness and the kurtosis
with the initial conditions based on the ZA and
3LPT  become about as much as $15 \%$ and $40 \%$
(at $z=2$), 
respectively. 
Furthermore, the difference between them based on
2LPT and 3LPT becomes also significant.
This means that the simulations with the initial
condition based on 2LPT does not provide 
the correct skewness because nonlinearity 
at this initial time is significant.
Without doubt, in such a case in which the initial
condition is imposed at 
$z \simeq 23$ like \cite{Crocce2006}, 
it is not enough to consider the initial condition
based on 2LPT and we can acquire much more accurate values
with the initial condition based on 3LPT.

\begin{figure}[tb]
\centerline{
\includegraphics[height=12cm]{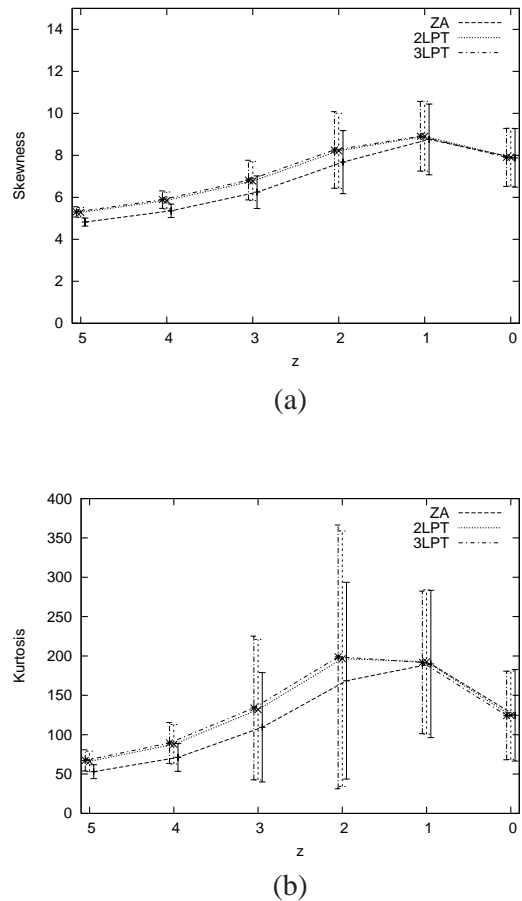}
}
\caption{
The non-Gaussianity of the density distribution
from N-body simulation 
($\delta_{max} (t_i)=0.5, R = 2 h^{-1}$ Mpc)
with different initial conditions. 
we also show the error bars.
(a) The skewness of the density distribution.
The difference between the skewness with the initial conditions
based on the ZA and 3LPT is less than  
$10 \%$.  Because of the suppressions of transients
at late time, this difference disappears at late time. 
Notice that there is no difference between the skewness with 
the initial conditions based on 2LPT and 3LPT 
since 2LPT provides almost exact skewness
for this initial time with small nonlinearity.
(b) The kurtosis of the density distribution.
The difference between the kurtosis with the initial conditions
based on the ZA and 3LPT is less than  
$25 \%$. Because of the suppressions of transients
at late time, this difference disappears at late time. 
Notice that the 
difference between the kurtosis with the initial conditions
based on 2LPT and 3LPT is less than $2\%$ and almost 
disappears at late time because transients with 
2LPT decrease much faster than the one with the ZA.
}
\label{fig:d05-NG}
\end{figure}

Next, we analyze the cases in which initial conditions are
imposed for smaller density fluctuations.
We reduce the maximum value
of the density fluctuation $\delta_{max} (t_i)$ 
at initial time to $0.5$, and $0.2$.
Decreasing $\delta_{max}$ will lead to less analytically evolved
solutions and more numerical evolution and hence is expected to
decrease the effect of transients which are due to 
analytical evolution.
For the case  $\delta_{max} (t_i)=0.5$, 
we can see that the difference between
the non-Gaussianity with the initial conditions based
on the ZA and 3LPT is much smaller than 
the case $\delta_{max} (t_i)=1.0$ (Fig.~\ref{fig:d05-NG}).
This is because for the smaller density fluctuation,
the higher order effect in LPT becomes less efficient
and the ZA is better approximation for the initial condition.
We can also see that the non-Gaussianity with the initial
condition based on the ZA approaches 
those with the initial condition based on 2LPT and 3LPT 
at late time. This means that transients disappear until
that.

It is worth noting here is that 
for $\delta_{max} (t_i)=0.5$,
there is no  difference between the skewness
with the initial conditions
based on 2LPT and 3LPT
since even 2LPT provides accurate skewness
for this initial time with small nonlinearity.
For the kurtosis, the difference with the 
initial conditions based on 2LPT and 3LPT
is also very small ( at most less than $2 \%$),
because transients with 2LPT decreases 
faster than that with the ZA.
This difference disappears until late time
as the transients vanishes at that time.

\begin{figure}[tb]
\centerline{
\includegraphics[height=12cm]{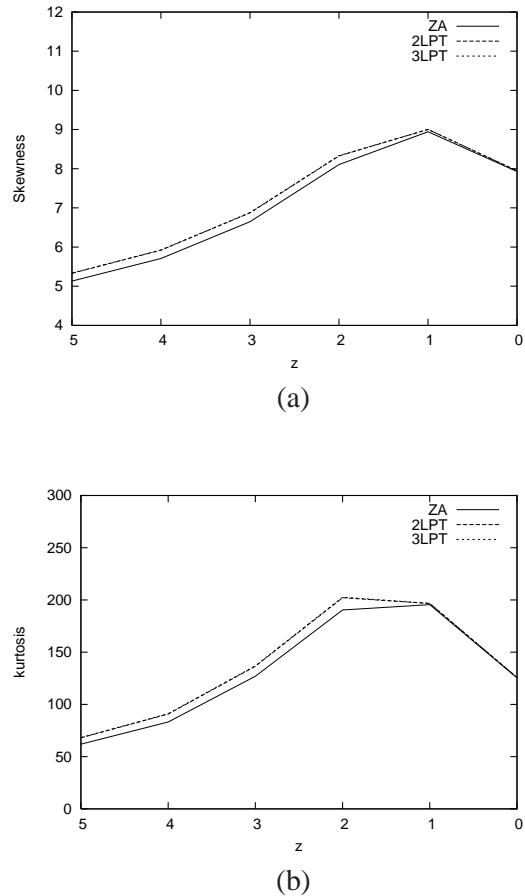}
}
\caption{The non-Gaussianity of the density distribution
from N-body simulation 
 ($\delta_{max} (t_i)=0.2, R = 2 h^{-1}$ Mpc)
with different initial conditions.
(a) The skewness of the density distribution.
There is no difference between the skewness with 
the initial conditions based on 2LPT and 3LPT 
since 2LPT provides almost exact skewness
for this initial time with small nonlinearity.
The difference between the skewness with the initial conditions
based on the ZA and 2LPT is less than  
$4 \%$ but remains until $z \sim 1$ because of transients.
(b) The kurtosis of the density distribution.
The difference between the kurtosis with the initial conditions
based on 2LPT and 3LPT become negligible because transients
with 2LPT have disappeared until $z\sim5$.
The difference between the kurtosiss 
with the initial conditions
based on the ZA and 2LPT is less than  
$10 \%$ but remains until $z \sim 1$ because of transients.
}
\label{fig:d02-NG}
\end{figure}

Furthermore, we examine the case $\delta_{max} (t_i)=0.2$
 (Figs.~\ref{fig:d02-NG}).
Both of the cases, the difference between the
non-Gaussianity
with the initial conditions
based on 2LPT and 3LPT almost disappears,
as is expected from the results of the case 
$\delta_{max} (t_i)=0.5$. Even for the kurtosis,
we cannot see the difference because before $z=5$,
transients with 2LPT completely disappear.

What is quantitatively important here is that
the difference between the skewness
and the kurtosis of the density distribution
with the initial conditions based on 
the ZA and 2LPT still survives for such a small value
of $\delta_{max} (t_i)=0.2$ until $z\sim1 $.
This means that it takes a long time for 
transients with the ZA to completely disappear.

By now, we only show the results after $z=5$,
even though we have actually done the numerical
calculations from about $z=20$.
For the reader who are interested in the thorough
evolution of the statistical quantities, we show
one example with $\delta_{max}(t_i)=0.5$
(Fig.~5).

\begin{figure}[tb]
\centerline{
\includegraphics[height=12cm]{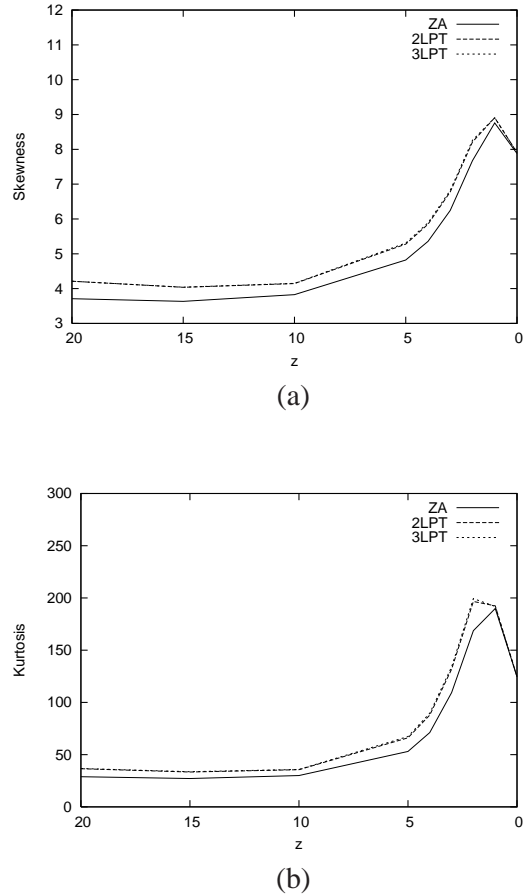}
}
\caption{
The non-Gaussianity of the density distribution
from N-body simulation 
($\delta_{max} (t_i)=0.5, R = 2 h^{-1}$ Mpc)
with different initial conditions. 
They are based on the same calculation as
in Fig.~\ref{fig:d05-NG}, but show the evolution
of the statistical quantities from $z=20$.
(a) The skewness of the density distribution.
(b) The kurtosis of the density distribution.
}
\label{fig:d05z-NG}
\end{figure}

To finish this section, it is worth commenting
on the error bars in the curves (Fig~{\ref{fig:d05-NG}}).
Although we only show error bars the 
$\delta_{max}(t_i) =0.5$ case, 
their values are almost 
independent of the initial conditions 
(ZA, 2LPT or 3LPT), hence hold for the other
simulations.
In Table~\ref{tab:er2} and Table~\ref{tab:er0},
we show the concrete values for the errors.

\begin{table}
\caption{\label{tab:er2} The error of the statistical
quantities at $z=2$ from N-body simulation 
($\delta_{max} (t_i)=0.5, R = 2 h^{-1}$ Mpc)
with different initial conditions. 
} 
\begin{tabular}{lccc}
\hline \hline
Initial condition  & Dispersion & Skewness & Kurtosis \\ \hline
ZA & $3.2 \% $ & $19.6 \%$ & $74.1 \%$ \\
2LPT & $3.4 \% $ & $21.8 \%$ & $82.6 \%$  \\
3LPT & $3.5 \% $ & $22.1 \%$ & $84.3 \%$ \\ \hline
\end{tabular}
\end{table}

\begin{table}
\caption{\label{tab:er0} The error of the statistical
quantities at $z=0$ from N-body simulation 
($\delta_{max} (t_i)=0.5, R = 2 h^{-1}$ Mpc)
with different initial conditions. 
} 
\begin{tabular}{lccc}
\hline \hline
Initial condition  & Dispersion & Skewness & Kurtosis \\ \hline
ZA & $6.9 \% $ & $17.8 \%$ & $46.6 \%$ \\
2LPT & $7.0 \% $ & $17.5 \%$ & $45.1 \%$  \\
3LPT & $7.0 \% $ & $17.5 \%$ & $45.2 \%$ \\ \hline
\end{tabular}
\end{table}


\section{Nonlinear effects in top-hat spherical
symmetric model}
 \label{sec:Nonlinear-effect}

According to the numerical calculation in the 
previous section, we see that even though
the difference between the
non-Gaussianity with the initial conditions based on
the ZA and 2LPT can not be removed well by
taking the initial time at sufficiently
high-$z$ era, it can be significantly decreased
between those based on 2LPT and 3LPT.
Especially, the difference between 
the kurtosis which initially includes transients 
with 2LPT has disappeared until $z\sim2$,
if we set the initial time for $N$-body
simulation around $z \sim 33$  (Table~\ref{tab:redshift}).

For the interpretation and further analytical
justification for the results, in this section,
we reconsider a simpler situation modeled by
a top-hat spherical symmetric collapse with
constant density for which both the exact solution
and Lagrangian perturbative expansion are
obtained analytically
\cite{Tatekawa04R,Munshi:1994zb,Sahni:1995rr}.
Another motivation for the symmetric 
collapse, is that it provides most severe
constraints for Lagrangian perturbation theory,
in general, as shown by  
Yoshisato \textit{et al.}~\cite{Yoshisato2006}.

In this section, we reset the definition of the scale factor $a$.
If such a shell is in the Einstein-de Sitter universe,
the equation of motion for a spherical shell is
given by
\begin{equation}
\frac{d}{dt} \left(a^2 \frac{dx}{dt}\right)
= -\frac{2 a^2 x}{9t^2}
\left[\left(\frac{x_0}{x}\right)^3 -1\right]\,,
\label{eom_th_shell}
\end{equation}
where $a(t) \propto t^{2/3}$ is a scale factor,
$x$ is a comoving radial coordinate and $x_0=x(t_0)$.
For the derivation of Eq.~(\ref{eom_th_shell}), see around
Eq.~(17) in \cite{Munshi:1994zb}.

Under the initial condition $\delta = a$ for $a \to 0$,
Eq.~(\ref{eom_th_shell}) can be integrated as,
\begin{equation}
\left( \frac{d R}{d a}\right)^2
=a \left( \frac{1}{R}- \frac{3}{5} \right)\,,
\label{exct_sol_ito_Ra}
\end{equation}
where $R(t)= a(t) x/x_0$ is a physical particle trajectory.
It is known that the exact solution for the 
spherical collapse (Eq.~(\ref{exct_sol_ito_Ra}))
can be parametrized as follows:
\begin{eqnarray}
R(\theta) = \frac{3}{10} (1 - \cos \theta),
\label{exct_sol_ito_theta1}\\
a(\theta) =\frac{3}{5} 
\left[\frac{3}{4} (\theta-\sin\theta)\right]^{2/3}\,.
\label{exct_sol_ito_theta2}
\end{eqnarray}
From Eqs.~(\ref{exct_sol_ito_theta1})  
and (\ref{exct_sol_ito_theta2}), we can also obtain
the density contrast  given by 
$\delta \equiv (x_0/x)^3 -1$ as
\cite{Tatekawa04R,Munshi:1994zb,Sahni:1995rr},
\begin{equation}
\delta (x) = \frac{9(\theta - \sin \theta)^2}
{2(1-\cos \theta)^3} -1\,.
\label{exct_sol_dens_cont}
\end{equation}
Together with 
Eqs.~(\ref{exct_sol_ito_theta2}),
(\ref{exct_sol_dens_cont})
and the definition of the density contrast,
we can express $R(t)$ exactly in terms of
the function $a(t)$.

Obviously, this exact expression can be 
expanded in terms of $a$ like 
\begin{equation}
R(t) = R_0 \left[ 1- \sum_{k=1} ^n 
(-1)^k C_k a^k\right]\,,
\label{lagrange_expansion_R}
\end{equation}
where $C_k$ are perturbative coefficients.
It can be shown that the first three terms 
of this expansion are given as
\begin{equation}
C_1 = \frac13,\;\;C_2=\frac{1}{21},\;\;
C_3 = \frac{23}{1701}\,.
\end{equation}

Since Eq.~(\ref{lagrange_expansion_R}) is 
an expansion with respect to the displacement
from the homogeneous distribution, this is nothing
but the Lagrangian perturbation.

On the other hand, for a general symmetric
spherical symmetric collapse, 
the density fluctuation in Eulerian picture at $r$
can be related with the perturbation variable in
Lagrangian picture as follows:
\begin{equation}
\delta(r) = \left( 1- \frac{\partial}{\partial r} s(r) 
\right)^{-3} -1\,,
\label{rel_delta_s}
\end{equation}
where $s(r)$ means the radial component of Lagrangian
displacement which can be expanded as
\begin{equation}
s(r) = s^{(1)} (r) + s^{(2)} (r) + s^{(3)} (r) + O(\varepsilon^4) \,.
\label{lagrange_expansion_s}
\end{equation}

Compared with Eq.~(\ref{lagrange_expansion_R})
and Eq.~(\ref{rel_delta_s}) together with 
the definition of the density contrast,
we can obtain the following relations:
\begin{equation}
\frac{\partial s}{\partial r} = 
\sum_{k=1} ^{n} (-1)^k C_k a^k
 \,,
\end{equation}
and applying the expansion given by 
Eq.~(\ref{lagrange_expansion_s}) this yields
\begin{eqnarray}
\frac{\partial }{\partial r} s^{(1)} (r)&=& \frac{1}{3}a,
\label{lpt__thc_s1}\\
\frac{\partial }{\partial r} s^{(2)} (r)&=& \frac{1}{21}a^2
 = \frac{3}{7} \left ( \frac{\partial}{\partial r}
  s^{(1)} (r) \right )^2 \,,
\label{lpt__thc_s2}\\
\frac{\partial }{\partial r} s^{(3)} (r)&=& \frac{23}{1701}a^3
 =\frac{23}{63} \left ( \frac{\partial}{\partial r}
  s^{(1)} (r) \right )^3
\,.
\label{lpt__thc_s3}
\end{eqnarray}
This spherical model forms  caustics at $a=3$ in the ZA.
Here we note that this is an ideal model,
and the value of $a$ which forms caustics is different from
the actual Universe.

%
%

In GRAFIC code, the initial conditions  are automatically 
set when the maximum density fluctuation 
at any lattice point becomes greater than some value.
If the point is approximated as 
the center of the spherical collapse, 
we can derive the relation between the density
fluctuation and the linear Lagrangian displacements
from Eq.~(\ref{rel_delta_s}).
\begin{equation}
\frac{\partial}{\partial r} s^{(1)} (r) =1-
 (1+ \delta (r))^{-1/3} \,.
\end{equation}
The linear Lagrangian displacements at this time becomes
\begin{eqnarray}
\frac{\partial}{\partial r} s^{(1)} (r) &\simeq& 0.206299 
\;\;\;\;{\rm for}\;\; \delta =1,\\
\frac{\partial}{\partial r} s^{(1)} (r) &\simeq& 0.126420
\;\;\;\;{\rm for}\;\; \delta =0.5,\\
\frac{\partial}{\partial r} s^{(1)} (r) &\simeq& 0.058964
\;\;\;\;{\rm for}\;\; \delta =0.2\,.
\end{eqnarray}
Therefore based on the discussions in this section,
we can calculate the effect of 2LPT and 3LPT at initial time
(Table~\ref{tab:nLPT}).

\begin{table}
\caption{\label{tab:nLPT} The effect of the second- and
third-order perturbation for the density fluctuation
evaluated at a given $\delta$
during the top-hat spherically symmetric collapse.
$\delta^{(i)}$ denotes that this quantity is calculated by
$i-$th order Lagrangian perturbation theory.
} 
\begin{tabular}{lcc}
\hline \hline
$\delta$ & $\delta^{(2)}/(\delta^{(1)}+\delta^{(2)})$ & 
$\delta^{(3)}/(\delta^{(1)}+\delta^{(2)}+\delta^{(3)})$\\ \hline
$1.0$ & $14.4 \%$ & $2.34 \%$ \\
$0.5$ & $7.17 \%$ & $0.73 \%$  \\
$0.2$ & $2.86 \%$ & $0.14 \%$ \\ \hline
\end{tabular}
\end{table}

Here we regard the  actual Universe with the above result.
Clearly, by regarding the top-hat spherical
collapse considered in this section as 
the generation of the overdense regions through 
gravitational instability in the Universe,
this analysis provides another explanation
for the results in the previous section.
For example, for the initial condition $\delta_{max} = 1$,
the effect of 3LPT has become significant,
while for $\delta_{max} = 0.5$ and 
$\delta_{max} = 0.2$  it is only below $1 \%$ and
eventually negligible.
On the other hand, the effect of 2LPT is still 
significant even we choose the initial condition
with smaller density fluctuation like $\delta_{max} = 0.2$.

\section{Summary}\label{sec:Summary}

Recently, observations have improved the precision
of cosmological constraints tremendously.
In order to compare theoretical predictions 
with observational results,
we should carry out detailed 
numerical simulations with high accuracy. 
In this context, 
in order to follow the nonlinear evolution,
one needs to consider carefully 
how to set initial conditions of 
$N$-body simulations very carefully.
In the standard method,
the initial conditions are set up by 
the Zel'dovich approximation (ZA)
in the quasi-nonlinear regime. 
The ZA has the nice property that by using linear 
perturbation theory allows one to describe accurately 
the quasi-nonlinear density field, in the case of
small non-linearity.

On the other hand, when constraining cosmological models,
non-Gaussianity of the density fluctuations generated 
by the nonlinear dynamics will become important.
It is well known that the ZA is not sufficient
to measure non-Gaussianity from higher-order
statistics, for example, the skewness and kurtosis.
The reason is the existence of so-called transients,
that is, the most dominant decaying mode arising from our
ignorance of the initial conditions. 
Until the effects of transients
have disappeared, we can not accurately reproduce 
higher-order quantities like the skewness and the kurtosis.

Actually, recently, 
Crocce, Pueblas, and Scoccimarro~\cite{Crocce2006}
analyzed the impact of transient on the 
second- and higher-order statistics 
quantitatively for $N$-body simulations
by comparing the initial conditions based on
the ZA and second-order Lagrangian perturbation theory
(2LPT). They show that even though there are also 
transients from 2LPT initial conditions,
the effects of transients are suppressed
compared to the ZA case.

In this paper, as a natural extension of 
\cite{Crocce2006}, in addition to the ZA and 2LPT,
we also analyze the non-Gaussianity with the initial
conditions based on third-order Lagrangian 
perturbation theory (3LPT). Since 3LPT initial 
conditions are expected to provide exact
results for the kurtosis
in the weakly nonlinear region, we can evaluate the impact
of transients from 2LPT initial conditions. 
The goal is to clarify the range where 
$N$-body simulations from 2LPT  initial conditions
are accurate.

When we set the initial condition of the maximal 
density contrast reaching unity $(\delta_{max} \sim 1)$
at $z \simeq 20$, differences in the predicted
non-Gaussianity between the three different initial
conditions (ZA, 2LPT and 3LPT) are readily apparent.
Since the accurate value
of the skewness is reproduced by 2LPT for the weakly
nonlinear regime, this disagreement is because
of the nonlinearity at this initial time.
Therefore, 2LPT initial conditions are not
sufficient to set up the initial condition for
$N$-body simulations.
There is also no guarantee that 3LPT initial conditions 
are accurate enough for precise determination
of cosmological parameters.

Next, to avoid the problem with the initial nonlinearity,
we set the initial conditions at $z \simeq 30$,
corresponding to the maximal density contrast
is about half, $(\delta_{max} \sim 0.5)$.
In this case, as is expected from the predictions of
2LPT in the weakly nonlinear region, 
we find no difference between the 2LPT and 3LPT results,
confirming that the 2LPT initial conditions produce 
accurate skewness.
For the kurtosis, the difference
between the initial conditions with 2LPT and 3LPT
is also very small ($2\%$). 
On the other hand, the difference of non-Gaussianity
with initial conditions based on the ZA and 3LPT is 
large ($10\%$ for the skewness and $25 \%$
for the kurtosis) until $z \sim 1$. 
This shows that transients from initial conditions
with 2LPT have less impact
than the ones with the ZA initial conditions.

This tendency is also obtained for the initial conditions
set at $z \simeq 80$  when the maximal density 
contrast is about two-tenths, $(\delta_{max} \sim 0.2)$.
From the numerical calculations for the kurtosis,
we can see that the effects of transients 
from 2LPT initial condition
have completely disappeared until $z \sim 5$.
However, there is still significant differences
in the predicted non-Gaussianity with initial conditions 
based on the ZA and 3LPT 
($4\%$ for the skewness and $10 \%$
for the kurtosis) until $z \sim 1$.

Therefore, for sufficiently early $z>30$ initial conditions,
the effects of transients on kurtosis from 2LPT
initial conditions become negligible until roughly 
$z \sim 3$, while those from the ZA initial conditions
survive until $z \sim 1$.
As long as one considers typical $N$-body simulations, which
start at $z \sim 50$, the predicted statistics are 
accurate enough up to the forth-order (kurtosis) 
using 2LPT initial conditions.

We further investigated our above results 
semi-analytically by considering the simpler situation
modeled by a top-hat spherically symmetric collapse
with constant density. We can show that in this situation,
the difference between the impact of 2LPT and 3LPT 
initial conditions
becomes almost negligible for the initial maximum 
density contrast  much less than
the half ($\delta_{max} \sim 0.5$), while
the difference between the impact of the ZA and 2LPT
initial conditions is still greater than $2\%$ 
for the initial maximum
density contrast about two-tenth 
($\delta_{max} \sim 0.2$).

Finally, we briefly mention the observational consequences
of our results. It is known that weak
lensing surveys can potentially provide us with
precision maps of the projected density up to redshifts
around 1
\cite{vanWaerbeke:2000rm,Bacon:2000sy,Wittman:2000tc,
Kaiser:2000if,Huterer:2004tr}.
Even though we need another step in obtaining 
the convergence field which can be written as the projection
of the matter density along the line of sight, the skewness
and kurtosis of the convergence field can be tested via 
weak lensing surveys. Our comparison of the initial
conditions should prove useful
extracting cosmological parameters from 
observational data.

\ack
SM is supported by JSPS.
We thank Masahiro Morikawa and Shoichi Yamada for useful comments.
We are grateful to Erik Reese for checking of 
English writing of this paper.


\section*{References}

\begin{thebibliography}{99}

\bibitem{Peebles}
Peebles P J E 1980 {\it The Large-Scale Structure of the Universe}
(Princeton: Princeton University Press)

\bibitem{Peacock}
Peacock J 1999 {\it Cosmological Physics}
 (Cambridge: Cambridge University Press)

\bibitem{Liddle}
Liddle A R and Lyth D H 2000
{\it Cosmological Inflation and Large-Scale Structure}
 (Cambridge: Cambridge University Press)

\bibitem{Coles}
Coles P and Lucchin F 2002
{\it Cosmology, The Origin and Evolution of 
Cosmic Structure} (Chichester: John Wiley)

\bibitem{Sahni-Coles}
Sahni V and Coles P 1995 
{\it Phys. Rep.} \textbf{262} 1

\bibitem{P3M}
Hockney W R and Eastwood W 1981 {\it Computer Simulation Using Particles}
 New York: (McGraw-Hill);
Bertschinger E and Gelb J M 1991 {\it Computers in Physics}
 \textbf{5} (2) 164

\bibitem{Bertschnger1998}
Bertschinger E 1998
{\it Ann. Rev. Astron. Astrophys.} \textbf{36} 599

\bibitem{Klypin83} Klypin A and Shandarin S F 1983
 {\it Mon. Not. R. Astron. Soc.}
 \textbf{204} 891

\bibitem{Efstathiou:1985re}
  Efstathiou G, Davis M, Frenk C S and White S D M 1985
  {\it Astrophys. J. Suppl. } \textbf{57} 241 

\bibitem{ZA} Zel'dovich Ya B 1970 {\it Astron. Astrophys.}
 \textbf{5} 84

\bibitem{Bernardeau02}
Bernardeau F, Colombi S, Gazta\~{n}aga E and Scoccimarro R 2002
{\it Phys. Rept.}  \textbf{367}, 1

\bibitem{Tatekawa04R} 
Tatekawa T 2005 {\it Recent Res. Devel. Astrophys.} \textbf{2} 1
[arXiv:astro-ph/0412025].


\bibitem{Juszkiewicz:1993uw}
  Juszkiewicz R, Bouchet F R and Colombi S 1993
  {\it Astrophys. J.}  \textbf{412} L9
  [arXiv:astro-ph/9306003].



\bibitem{Juszkiewicz:1993hm}
  Juszkiewicz R, Weinberg D H, Amsterdamski P, Chodorowski M and Bouchet F
  1995
  {\it Astrophys. J.}  \textbf{442} 39



\bibitem{Baugh:1995hv}
  Baugh C M, Gaztanaga E and Efstathiou G 1995
  {\it Mon. Not. Roy. Astron. Soc.}  \textbf{274} 1049



\bibitem{Scoccimarro:1997gr}
  Scoccimarro R 1998
  {\it Mon. Not. Roy. Astron. Soc. }  \textbf{299} 1097
  [arXiv:astro-ph/9711187].


\bibitem{Valageas:2001}
  Valageas P 2002
  {\it Astron. Astrophys.} \textbf{385} 761



\bibitem{Crocce2006} Crocce M, Pueblas S and
 Scoccimarro R 2006
  {\it Mon. Not. R. Astron. Soc.}
  \textbf{373} 369

\bibitem{Bouchet92}
Bouchet F R, Juszkiewicz R, Colombi S and Pellat R 1992
  {\it Astrophys. J.} \textbf{394} L5

\bibitem{Buchert93}
Buchert T and Ehlers J 1993
  {\it Mon. Not. R. Astron. Soc.} \textbf{264} 375

\bibitem{Munshi:1994zb}
  Munshi D, Sahni V and Starobinsky A A 1994
  {\it Astrophys. J. } \textbf{436} 517
  [arXiv:astro-ph/9402065].

\bibitem{Buchert94}
  Buchert T 1994 {\it Mon. Not. R. Astron. Soc.} \textbf{267} 811

\bibitem{Bouchet95}
  Bouchet F R, Colombi S, Hivon E, and Juszkiewicz R,
  {\it Astron. Astrophys.} \textbf{296} 575

\bibitem{Catelan95} Catelan P 1995
  {\it  Mon. Not. R. Astron. Soc.} \textbf{276} 115

\bibitem{Sahni:1995rr}
  Sahni V and Shandarin S F 1996
  {\it Mon. Not. Roy. Astron. Soc. }  \textbf{282} 641
  [arXiv:astro-ph/9510142].


\bibitem{transverse}
Strictly speaking, even if we ignore
pressure of fluid, it is shown that 
third-order transverse mode appears~\cite{Sasaki98},
in principle.
In generic cases, however, the effect of the third-order
transverse mode is quite small quantitatively.
Furthermore, if we consider special situations 
when the fluid affects some kind of pressure gradient,
transverse mode appears from the 
second-order~\cite{Pressure}.

\bibitem{Sasaki98}
  Sasaki M and Kasai M 1998
  {\it Prog. Theor. Phys.} \textbf{99} 585

\bibitem{Pressure}
  Morita M and Tatekawa T 2001
  {\it Mon. Not. R. Astron. Soc.} \textbf{328} 815;
  Tatekawa T~{\it et al.} 2002
  {\it Phys. Rev. D} \textbf{66} 064014;
  Tatekawa T 2005
  {\it Phys. Rev. D} \textbf{72} 024005

\bibitem{COSMICS}
  Ma C P and Bertschinger E 1995 {\it Astrophys. J.} \textbf{455} 7

\bibitem{WMAP} Spergel D N {\it et al.}, astro-ph/0603449.


\bibitem{Yoshisato2006}
  Yoshisato A, Morikawa M, Gouda N and Mouri H 2006
  {\it Astrophys. J.} \textbf{637} 555
  [arXiv:astro-ph/0510107].

\bibitem{vanWaerbeke:2000rm}
  van Waerbeke L \textit{et al.} 2000
  {\it Astron. Astrophys.}  \textbf{358} 30
  [arXiv:astro-ph/0002500].

\bibitem{Bacon:2000sy}
  Bacon D J, Refregier A R and Ellis R S 2000
  \textit{Mon. Not. Roy. Astron. Soc.}  \textbf{318} 625
  [arXiv:astro-ph/0003008].

\bibitem{Wittman:2000tc}
  Wittman D M, Tyson J A, Kirkman D, Dell'Antonio I and Bernstein G 2000
  \textit{Nature} \textbf{405} 143
  [arXiv:astro-ph/0003014].

\bibitem{Kaiser:2000if}
  Kaiser N, Wilson G and Luppino G A,
  [arXiv:astro-ph/0003338].

\bibitem{Huterer:2004tr}
  Huterer D and Takada M 2005
  {\it Astropart. Phys.} \textbf{23} 369
  [arXiv:astro-ph/0412142].



\end{thebibliography}
\end{document}